\documentclass[journal]{IEEEtran}
\usepackage{cite}

\ifCLASSINFOpdf
   \usepackage[pdftex]{graphicx}
\else
   \usepackage[dvips]{graphicx}
\fi

\begin{document}
\title{High energy resolution hard X-ray and gamma-ray imagers using CdTe diode devices}
\author{Shin~Watanabe, Shin-nosuke~Ishikawa, Hiroyuki~Aono, Shin'ichiro~Takeda,
Hirokazu~Odaka, Motohide~Kokubun, Tadayuki~Takahashi, Kazuhiro~Nakazawa, Hiroyasu~Tajima,
Mitsunobu Onishi, Yoshikatsu Kuroda 

\thanks{S.~Watanabe, S.~Ishikawa, H.~Aono, S.~Takeda, H.~Odaka and T.~Takahashi
are with Institute of Space and Astronautical Science, Japan Aerospace Exploration Agency, Sagamihara, Kanagawa, Japan, 
and also with Department of Physics, University of Tokyo, Bunkyo, Tokyo, Japan.} 
\thanks{M.~Kokubun is with Institute of Space and Astronautical Science, Japan Aerospace Exploration Agency.} 
\thanks{K.~Nakazawa is with Department of Physics, University of Tokyo.} 
\thanks{H.~Tajima is with Stanford Linear Accelerator Center, Menlo Park, CA, USA.} 
\thanks{M.~Onishi and Y.~Kuroda are with Nagoya Guidance and Propulsion Systems Works, Mitsubishi Heavy Industry Ltd., Komaki, Aichi, Japan.} 
\thanks{Manuscript accepted Oct 12, 2008}}

\maketitle

\begin{abstract}
We developed CdTe double-sided strip detectors (DSDs or cross strip detectors) 
and evaluated their spectral and imaging performance for hard X-rays and gamma-rays.
Though the double-sided strip configuration is suitable for imagers with a
fine position resolution and a large detection area, CdTe diode DSDs with indium (In) anodes
have yet to be realized due to the difficulty posed by the segmented In anodes.
CdTe diode devices with aluminum (Al) anodes were recently established, followed by 
a CdTe device in which the Al anodes could be segmented into strips.
We developed CdTe double-sided strip devices having Pt cathode strips 
and Al anode strips, and assembled prototype CdTe DSDs. 
These prototypes have a strip pitch of 400~$\mu$m.
Signals from the strips are processed with analog 
ASICs (application specific integrated circuits). We have successfully
performed gamma-ray imaging spectroscopy with a position resolution of 400~$\mu$m.
Energy resolution of 1.8~keV (FWHM: full width at half maximum) was obtained at
59.54~keV. Moreover, the possibility of improved spectral performance
by utilizing the energy information of both side strips was demonstrated.
We designed and fabricated a new analog ASIC, VA32TA6, 
for the readout of semiconductor detectors, which is also suitable for DSDs.
A new feature of the ASIC is its internal ADC function.
We confirmed this function and good noise performance that reaches 
an equivalent noise charge of 110~e$^-$ under the condition of 
3--4~pF input capacitance.  
\end{abstract}


\IEEEpeerreviewmaketitle

\section{Introduction}
\IEEEPARstart{H}{ard} X-ray and gamma-ray imaging spectrometers with good spatial and energy
resolutions are desired for medical, industrial and astrophysical applications.
Cadmium telluride (CdTe) and cadmium zinc telluride (CZT) are very promising 
materials for hard X-ray and gamma-ray imaging spectrometers,
given their high detection efficiency comparable to that of NaI scintillators 
and good energy resolution comparable to that of Ge detectors.
Although both CdTe and CZT are vulnerable to energy resolution and
peak detection efficiency being degraded due to their poor charge transport properties, 
several techniques have been developed to maintain good spectral 
performance \cite{takahashiwatanabe01, limousin}.

\begin{figure}[!t]
\centering
\includegraphics{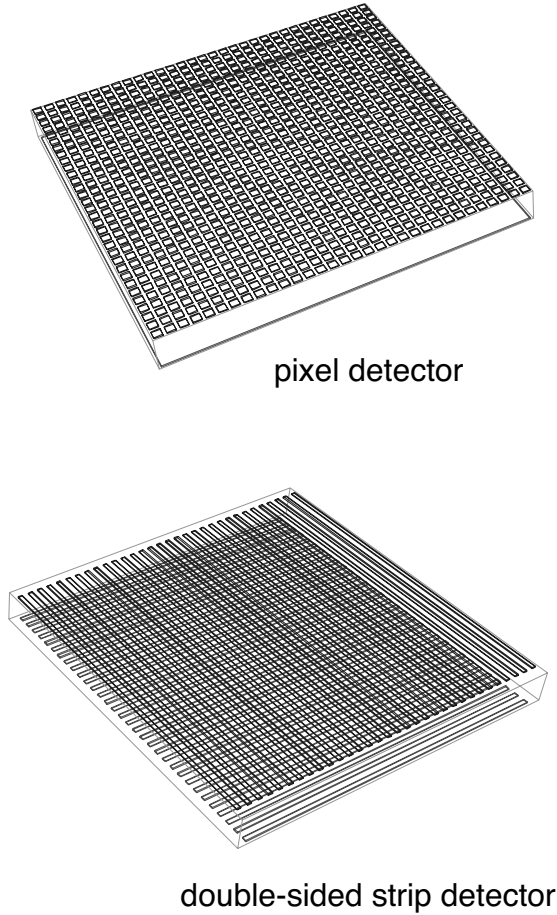}
\caption{Pixel detector and double-sided strip detector.}
\label{fig:pixeldssd}
\end{figure}

CdTe/CZT semiconductor detectors require 
segmented readout electrodes in order to obtain position information.
For imagers having a fine position resolution, there are two types
of detector configurations: pixel detectors and double-sided strip detectors 
(Fig.~\ref{fig:pixeldssd}).

A pixel detector has a number of small pixel electrodes on one side.
The signal from each pixel is processed with each charge-sensitive amplifier.
Since the leakage current and the detector capacitance become very small,  
ideal spectral performance is consequently possible. However, an extremely large
number of readout channels are needed for a fine position resolution and/or  
large detection area. Additionally, a two-dimensional readout ASIC is essential
for fine pitch pixels \cite{cztpixel1, cztpixel2, cdtepixel1}.

A double-sided strip detector (DSD or cross strip detector) realizes a fine position resolution 
and large detection area with a relatively small number of readout channels.
The DSD has orthogonal strips implemented on both its sides.
By reading out the signal from both side strips coincidentally, it is possible to obtain
information on the position and energy of the X-ray/gamma-ray photons \cite{cztstrip1, cztstrip2}. 
The number of the readout channels is proportional to $2N$ (where $N$ denotes the 
number of segments per coordinate) for DSDs, instead of $N^2$ for pixel detectors.
Therefore, for a larger $N$ (meaning a finer position resolution and/or larger area),
the DSD has an advantage in terms of readout channels. 
Moreover, one-dimensional ASICs that are easier to implement 
and more common than two-dimensional ones are also applicable to DSDs.

Our development is aimed at hard X-ray imaging detectors for the ASTRO-H (NeXT) project 
that plans to launch Japan's 6th X-ray astronomy satellite in 2013. 
Among the major objectives is achieving high sensitivity observation with focusing and imaging 
capabilities in the 5--80~keV energy region.
The ASTRO-H satellite will carry two hard X-ray telescopes that feature  
multilayer supermirrors assembled in grazing incident X-ray telescopes. 
Hard X-ray Imagers (the focal plane detectors of hard X-ray telescopes) require 
an energy resolution of $\sim$~1~keV (FWHM: full width at half maximum), energy coverage of 5--80~keV,
a sub-mm (250--500~$\mu$m) position resolution, detection area of 2--3~cm, and timing resolution of $\sim$~1~$\mu$s
for high sensitivity observation.
To achieve these goals, we have developed CdTe diode DSDs as a primary choice for Hard X-ray Imagers.  
In this paper, we describe our development of CdTe DSDs, and report the results of
the CdTe DSD prototypes. We also report on our newly developed ASIC (VA32TA6) for semiconductor detector readout.

\section{Prototypes of CdTe DSDs}
\subsection{CdTe Diode Device with Double-Sided Strips}
Thin CdTe diode devices that utilize indium (In) as the 
anode electrodes on p-type CdTe wafers manufactured by ACRORAD (Japan) and platinum (Pt) as the cathodes 
have been established and offer good spectral performance \cite{takahashiwatanabe01}. 
A high Schottky barrier formed on the In/p-CdTe interface enables
use of the detector as a diode. Using this type of detector with a thickness 
of 0.5~mm at applied bias voltages
as high as 1~kV makes it possible to overcome the poor charge transport properties of CdTe.
However, double-sided strip detectors with In anodes cannot be realized 
since it is difficult to segment the In electrode into strips.

Aluminum (Al) has recently been found to be a good alternative electrode material
to In \cite{alanode1, alanode2}. In addition to low leakage currents and high energy 
resolutions comparable to those of In/CdTe/Pt detectors, 
Al/CdTe/Pt detectors also offer the advantage of allowing Al anodes to 
be divided into pixels or strips. 
CdTe diode detectors with segmented Al anodes have been established
by constructing and testing Al-pad/CdTe/Pt or Al-pixel/CdTe/Pt type detectors.
Good spectral performance ($\Delta E$~$\sim$~1~keV(FWHM) at 60~keV) has been achieved \cite{alpixel,alpixel2}.

By adopting segmented Al anodes and Pt cathodes,
we can obtain a CdTe diode double-sided strip device.
For the prototype detectors, we fabricated two types of the CdTe devices:
the 2.6~cm CdTe device and 1.3~cm CdTe device.
The 2.6~cm CdTe device is 2.6~cm~$\times$~2.6~cm in size, and
500~$\mu$m thick. On each side 64 strips are formed, with a strip pitch
of 400~$\mu$m. Moreover, 350~$\mu$m strip electrodes and 50~$\mu$m gaps are alternately formed.
The 1.3~cm CdTe device is 1.3~cm~$\times$~1.3~cm in size, and
500~$\mu$m thick. On each side 32 strips are formed. The strip pitch and
electrode configuration are the same as those of the 2.6~cm CdTe DSD.
Both devices have guard-ring electrodes on both sides to reduce 
leakage current of the strips.

\subsection{Configurations of Prototypes}
\begin{figure}[!t]
\centering
\includegraphics{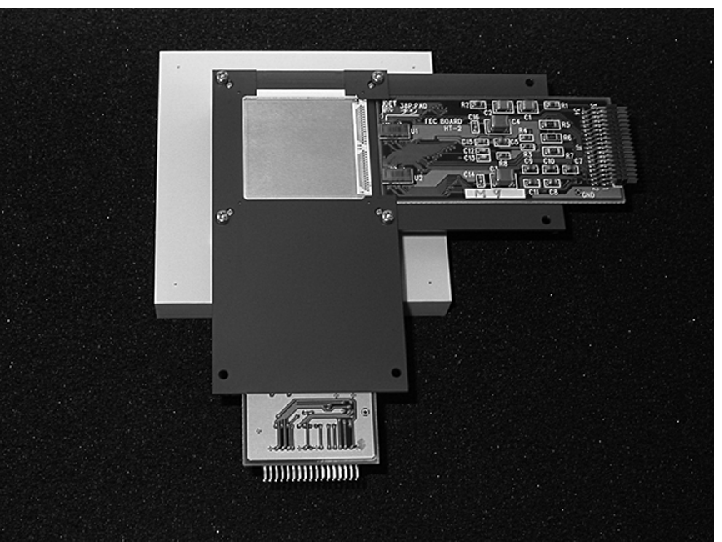}

(a)

\medskip
\includegraphics{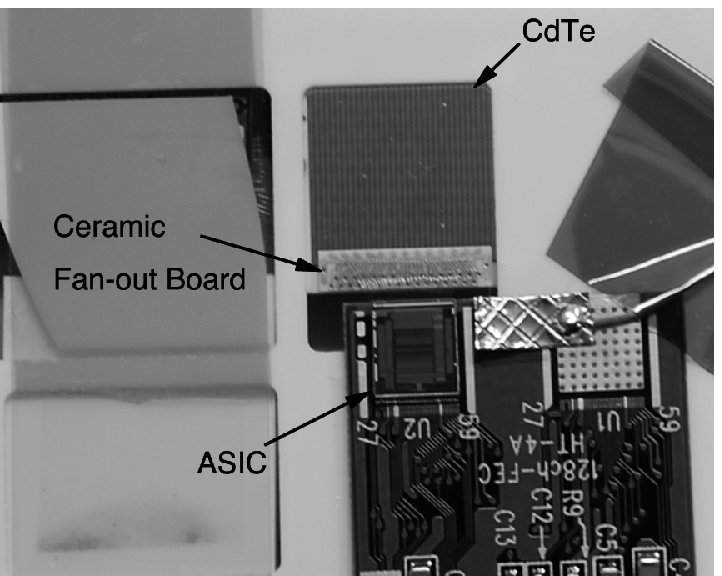}

(b)
\caption{Prototypes of the CdTe DSDs. 
(a): The 2.6~cm CdTe DSD. It is 2.6~cm~$\times$~2.6~cm in size and,
500~$\mu$m thick. The strip pitch is 400~$\mu$m. On each side 64 strips are
formed, and two VA32TAs are used for readout on each side.
(b): The 1.3~cm CdTe DSD. It is 1.3~cm~$\times$~1.3~cm in size, and
500~$\mu$m thick. The strip pitch is 400~$\mu$m. On each side 32 strips are
formed, and one VA64TA2 is used for readout on each side.}
\label{fig:prototype}
\end{figure}

We used the CdTe double-sided strip devices described above to assemble two types of prototype CdTe DSDs:
the 2.6~cm CdTe DSD and the 1.3~cm CdTe DSD. 
Fig.~\ref{fig:prototype} shows the prototype detectors. 
We employed VA32TAs \cite{Tajima_ieee04} and VA64TA2s \cite{ttanaka2006} 
that we jointly developed with GM-IDEAS 
to read out the 2.6~cm and the 1.3~cm CdTe DSDs, respectively. 

Given the soft and fragile characteristics of CdTe material, we have yet to establish a reliable 
technique of wire-bonding on the CdTe surface. Therefore, it is difficult to interconnect 
the CdTe strip electrodes and the readout ASIC input pads by using wire-bonding as applied to Si DSDs \cite{Tajima03, nimatakeda}.

To overcome this problem, we adopted the In/Au stud bump bonding technique and used the Al$_2$O$_3$ ceramic
fan-out board with through-holes. First, the CdTe strip electrodes are connected via In/Au stud bump bonding to the Al$_2$O$_3$ ceramic fan-out
board that we developed and established for CdTe-Pixel/Pad detectors \cite{ieee2001takahasi}. 
This ceramic fan-out board has
through-holes to interconnect the electrodes on both sides of the fan-out board. Then, wire-bonding can be done from the ASIC input pads
to the electrodes on the ceramic board. We have connected readout ASICs on both sides of the CdTe DSDs in this manner. 
A more detailed procedure is described in elsewhere \cite{cdtedsd1}.

\begin{figure}[!t]
\centering
\includegraphics[width=0.49\textwidth]{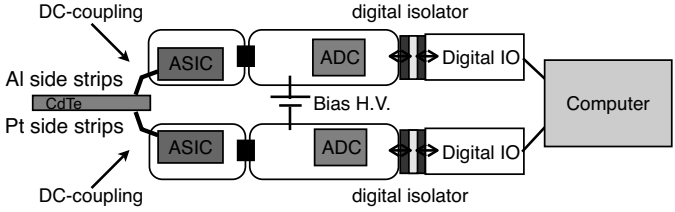}
\caption{Readout system of the CdTe DSDs.}
\label{fig:readout}
\end{figure}

Fig.~\ref{fig:readout} shows a block diagram illustrating the overall readout system. 
This system is virtually the same as the scheme implemented for the readout of Si DSDs \cite{Tajima03,nimatakeda}. 
The readout ASICs and strip electrodes are connected
with DC-coupling for both sides. The detector bias voltage is supplied by applying a high voltage between local grounds
of the Al anodes and Pt cathodes. 

\subsection{Imaging and Spectral Performance}
\begin{figure}[!t]
\includegraphics[width=0.24\textwidth]{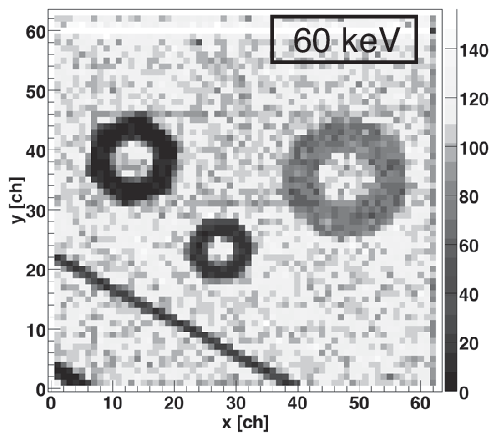}
\includegraphics[width=0.24\textwidth]{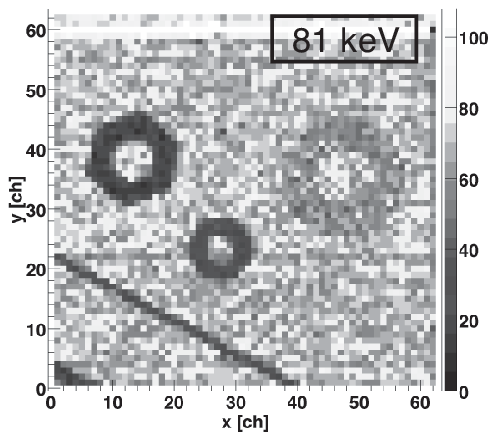}

\vspace{2mm}

\includegraphics[width=0.24\textwidth]{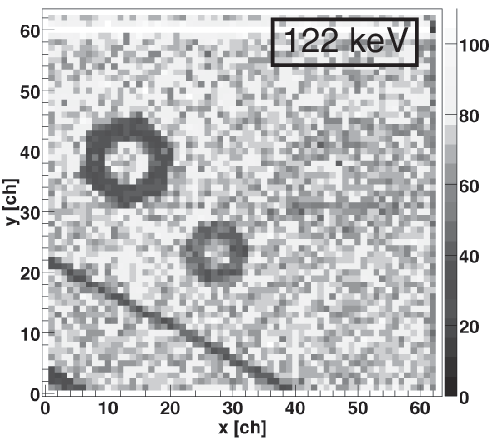}
\hspace{0.02\textwidth}
\includegraphics[width=0.2\textwidth]{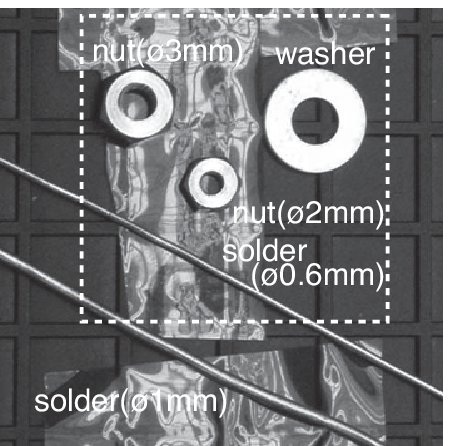}
\caption{Shadow images obtained with the 2.6~cm CdTe DSD and a photograph of the target. 
The energies of the gamma rays are 60~keV ($^{241}$Am), 81~keV ($^{133}$Ba) and 122~keV  ($^{57}$Co).
The pixel size in the images corresponds to a strip pitch of 400~$\mu$m.}
\label{fig:shadowImage}
\end{figure}

In order to demonstrate the imaging capability of the test devices, we took shadow images of metal objects
with various gamma rays from radioisotopes such as $^{241}$Am, $^{133}$Ba and $^{57}$Co. Fig.~\ref{fig:shadowImage}
shows the shadow images obtained with the 2.6~cm CdTe DSD and the target.  
The hole of a 2-mm nut and solder 0.6~mm diameter can be clearly seen. It can also be seen that
a thin washer becomes transparent as the energy of gamma rays becomes higher.
However, the energy resolution obtained with the 2.6~cm CdTe DSD did not match that obtained with 
pad-type Al anode detectors. FWHM energy resolutions of 2.6~keV and 6.2~keV at 59.54~keV were obtained for the spectra from Pt 
strips and Al strips, respectively, under 500~V of bias voltage and temperature of $-$20$^\circ$C.

For a detailed study of spectral performance, we tested a 1.3~cm CdTe DSD.
The smaller capacitance and lower leakage current afforded by smaller detector size should
lead us to better noise performance. The spectral performance of VA64TA2 for negative signals from the Al strips
has also been improved compared with VA32TA used in the 2.6~cm CdTe DSD.

\begin{figure}[!t]
\centering
\includegraphics[width=0.4\textwidth]{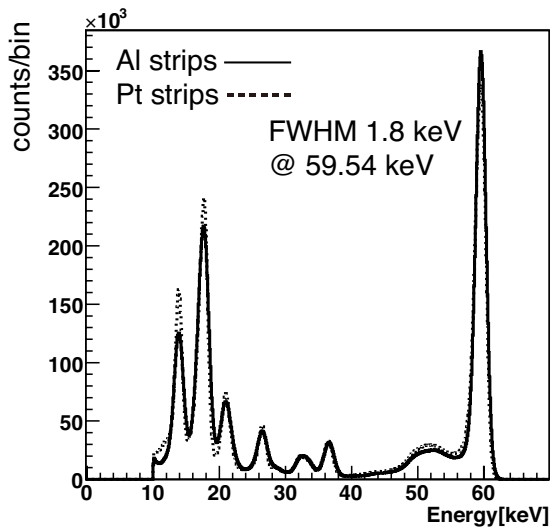}
\caption{$^{241}$Am spectra obtained with the 1.3~cm CdTe DSD. 
The solid and dotted lines indicate spectra from the Al and Pt strips, respectively.
A bias voltage of 500~V was applied at operating temperature of $-$20$^\circ$C.
An energy resolution of 1.8~keV(FWHM) at 59.54~keV was obtained for both sides.}
\label{fig:spec}
\end{figure}

Fig.~\ref{fig:spec} show the spectra obtained with the 1.3~cm CdTe DSD. Under 
operating conditions of $-$20$^\circ$C temperature and 500~V of applied bias voltage,  
the total leakage current from all strips and guard-rings was 5~nA. 
The trigger energy threshold can be set to about 10~keV.
An energy resolution of 1.8~keV (FWHM) at 59.54~keV was obtained for both the Pt and Al sides.
The spectral performance for the Al and Pt sides was confirmed as being equivalent in the CdTe DSDs.

\begin{figure}[!t]
\centering
\includegraphics[width=0.4\textwidth]{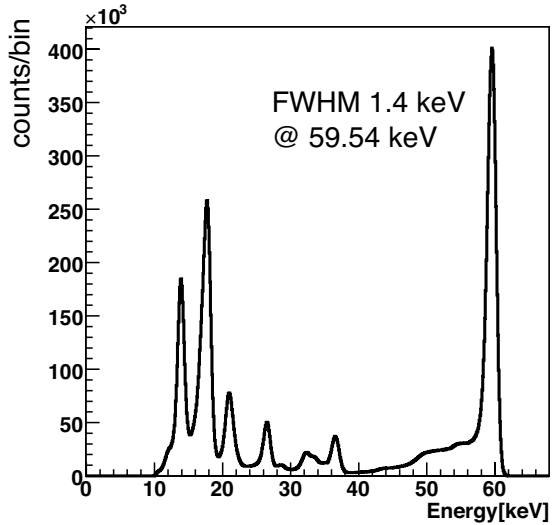}
\caption{$^{241}$Am spectra obtained with the 1.3~cm CdTe DSD. The average of the both sides' energy information was used.
The energy resolution was improved to 1.4~keV(FWHM) at 59.54~keV.
}
\label{fig:spec2}
\end{figure}

By collectively using the energy information on both sides, improved
spectral performance could be expected if each noise component was independent.
We created a spectrum by filling the average of the both sides' energy information, as 
shown in Fig.~\ref{fig:spec2}.
This spectrum shows improved energy resolution. Thus,
an energy resolution of 1.4~keV (FWHM) was successfully obtained.

\begin{figure}[!t]
\centering
\includegraphics[width=0.45\textwidth]{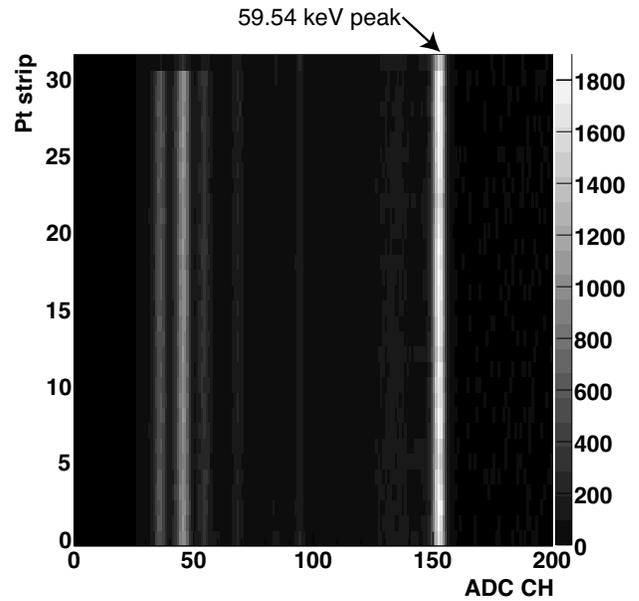}
\caption{$^{241}$Am spectra obtained with one Al strip of the 1.3~cm CdTe DSD. The X and Y axes correspond to
the Al strip's pulse height in the ADC channel and the hit strip on the Pt side, respectively.
}
\label{fig:uniformity}
\end{figure}

For imagers, uniform spectral performance in each detector is important.
In the double-sided strip configuration, we can examine a kind of uniformity by sorting 
the events of one strip based on the position information obtained from the other side's strips.
Since only one channel of the ASIC is used for the readout, we can obtain the position 
dependence without the effect of readout variation. Fig.~\ref{fig:uniformity} shows a $^{241}$Am gamma-ray spectrum
from one strip on the Al side. The Y axis corresponds to the hit strip on the Pt side. 
It can be seen that the spectrum is constant with the position.
The peak position and energy resolution for 59.54~keV gamma rays were stable within 0.2\% and 12\%, respectively.

\section{Brand New Readout ASIC, VA32TA6}
\begin{figure}[!t]
\centering
\includegraphics[width=2in]{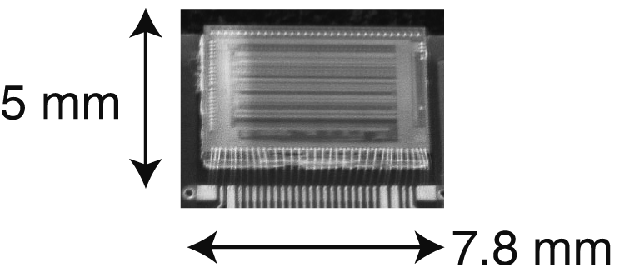}
\includegraphics[width=0.49\textwidth]{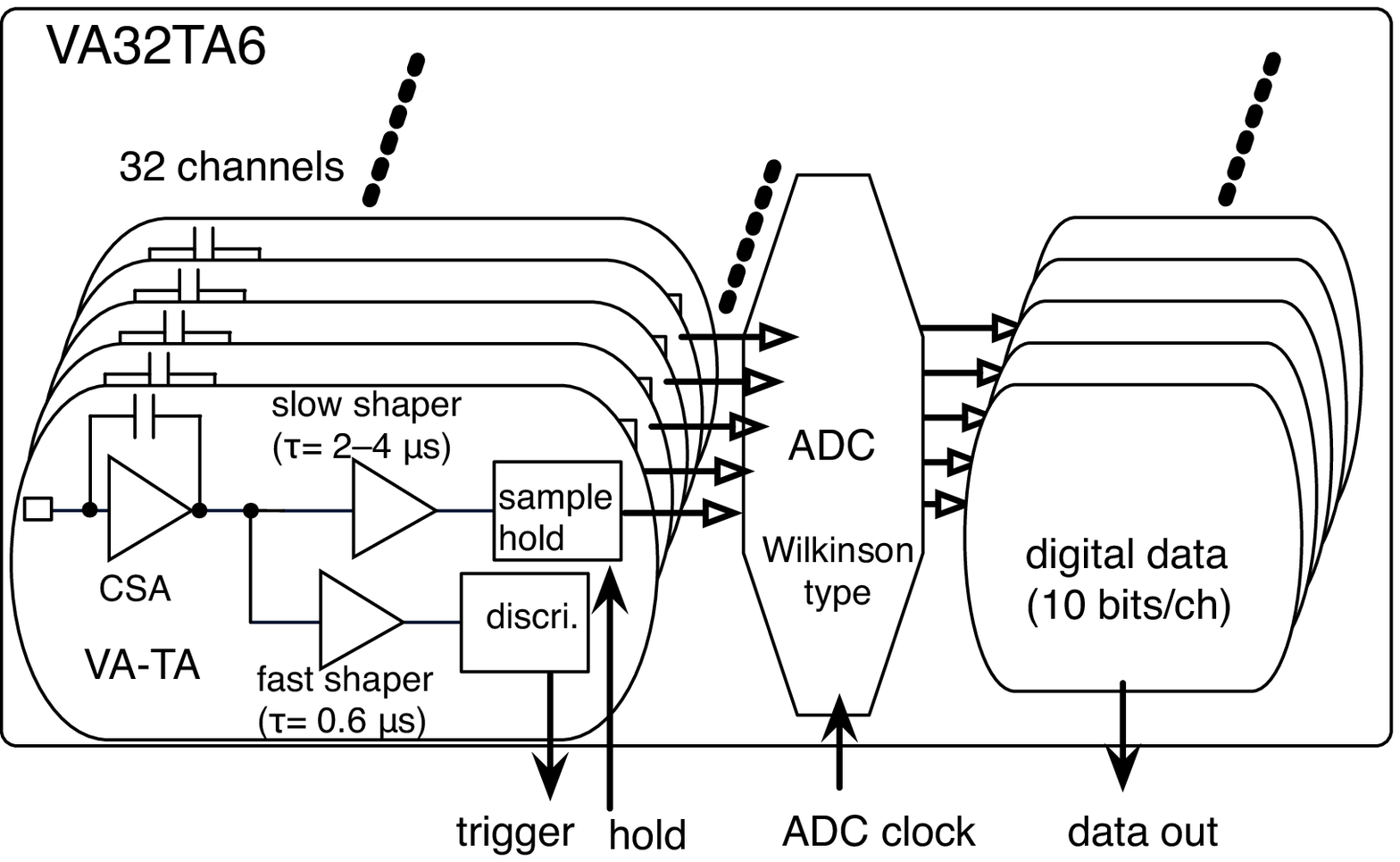}
\caption{Photograph of VA32TA6 and a schematic diagram.}
\label{fig:va32ta6}
\end{figure}

We have been developing readout ASICs for our applications in collaboration with GM-IDEAS.
Based on past developments, we designed and fabricated a new readout ASIC, VA32TA6.
Fig.~\ref{fig:va32ta6} shows a photograph of VA32TA6 and a schematic diagram of the ASIC.
VA32TA6 is fabricated using AMS 0.35~$\mu$m technology with an epitaxial layer. 
The die is 5.0~mm~$\times$~7.8~mm in size and 725~$\mu$m thick. 
The front-end part of VA32TA6 is based on VA64TA1 \cite{ttanaka2006}. It has 32 channels of circuits,
including a charge-sensitive amplifier, slow CR-RC shaper and sample/hold (VA section), and fast CR-RC shaper and
discriminator chain (TA section). The shaping time of the slow shaper is variable from 2 to 4~$\mu$s: that
of the fast shaper is 600~ns. A detailed description of this VA-TA architecture is given in other references \cite{Tajima03,Tajima_ieee04}.
The measured power consumption on average was 16.5~mW per chip, corresponding to 
0.5~mW per channel by considering a simple division of channel numbers.

The main new feature of VA32TA6 is the inclusion of on-chip ADC circuitry.
VA32TA6 implements a Wilkinson-type ADC that digitizes the 32 sampled pulse heights of the
slow shapers in parallel. 
The ramp speed is adjustable via the internal DAC. The ADC clock's maximum speed is 10~MHz. 
Each channel has a 10-bit digital counter. The maximum dead time due to AD conversion is
$\sim$~100~$\mu$s under the 10~MHz ADC clock.
The ADC circuitry is equipped with a global Common Mode Detection Unit.
This unit latches the ADC counter value when a certain number of channel discriminators have fired.
For VA32TA6, the threshold of the Common Mode Detection Unit has one default setting (16 channels).
The common mode value is stored in a separate counter and
output as part of the data stream.

To evaluate the functions and performance of the ASIC,
we constructed CdTe pad detectors for connection to VA32TA6. Each pad is 
1.35~mm~$\times$~1.35~mm in size. A total of 32~pads (in 8~$\times$~8~=~64 pads) are connected to
the ASIC. The pad capacitance is expected to be 3--4~pF per channel.
We used two types of CdTe pad detectors. 
One is an In/CdTe/Pt-pad type detector \cite{watanabenima2007}, 
in which VA32TA6 is connected to Pt cathode pads, 
for processing positive charge signals.
The other is an Al-pad/CdTe/Pt type detector \cite{alpixel}.
The ASIC in this detector is connected to Al anode pads
for processing negative charge signals.

\begin{figure}[!t]
\centering
\includegraphics[width=0.4\textwidth]{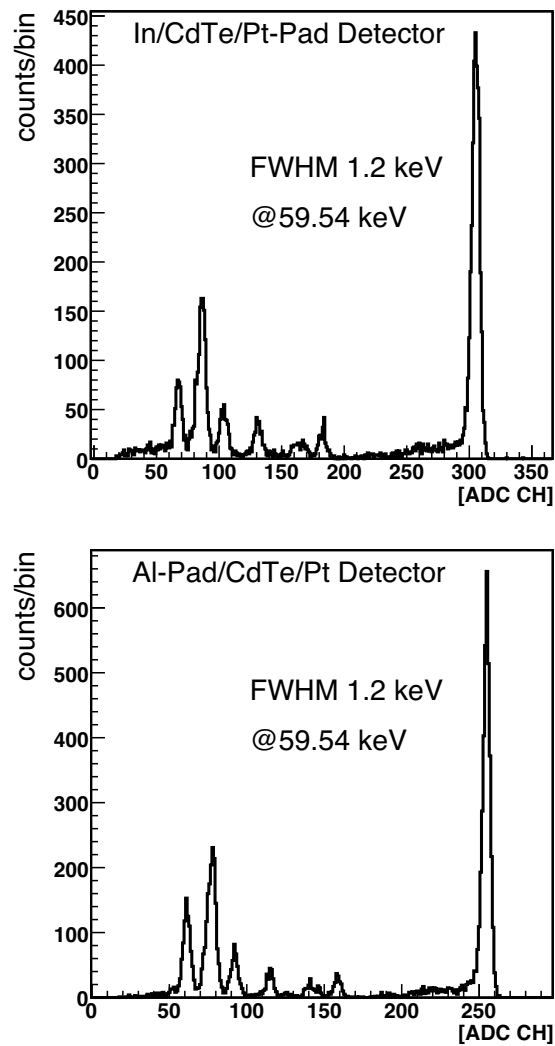}
\caption{$^{241}$Am gamma-ray spectra obtained with CdTe pad detectors using VA32TA6. 
At an operating temperature of $-$20$^\circ$C, bias voltages of 600~V
and 400~V were applied for the In/CdTe/Pt-pad detector and Al-pad/CdTe/Pt detector, respectively.
The pedestal level was corrected, 
and the common mode was subtracted by using the data from the Common Mode Detection Unit.
An energy resolution of 1.2~keV (FWHM) at 59.54~keV was achieved.}
\label{fig:va32ta6spec}
\end{figure}

By applying a detector bias voltage and irradiating the detector with X-rays/gamma rays emitted by a radioisotope, 
we have successfully made VA32TA6s work and have obtained digital pulse height data.
The ADC circuitry works properly with clock speed up to 10~MHz.
Fig~\ref{fig:va32ta6spec} shows the obtained spectra from the CdTe pad detectors.
These spectra are obtained from a single channel of the pad detectors. The pedestal level is corrected, 
and the common mode is subtracted by using the data from the Common Mode Detection Unit.
An energy resolution of 1.2~keV (FWHM) at 59.54~keV was achieved for both pad detectors. 
This energy resolution in CdTe corresponds to 110~e$^-$ in ENC (equivalent noise charge).
This spectral performance is comparable with that of VA64TA1 \cite{ttanaka2006}.

\begin{figure}[!t]
\centering
\includegraphics[width=0.45\textwidth]{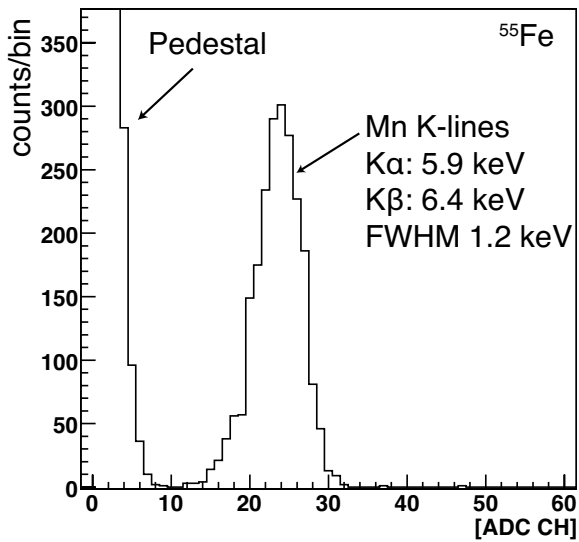}
\caption{$^{55}$Fe X-ray spectrum obtained with the Al-pad/CdTe/Pt detector 
using VA32TA6. At operating temperature of $-$20$^\circ$C, 400~V of  
bias voltage was applied. The pedestal level was corrected,
and the common mode was subtracted by using the data from the Common Mode Detection Unit.}
\label{fig:va32ta6spec2}
\end{figure}

For our astrophysical application, it is important to achieve as low an energy threshold
as possible. Fig~\ref{fig:va32ta6spec2} demonstrates the low energy threshold.
This is an X-ray spectrum from $^{55}$Fe obtained with one channel of the Al-pad/CdTe/Pt detector.
The Mn K X-ray line combining a 5.9~keV K$\alpha$ and a 6.4~keV K$\beta$ was
detected and clearly resolved. The energy threshold can be set as low as 4~keV, which
satisfies the energy coverage goal of Hard X-ray Imagers used in the ASTRO-H (NeXT) project.

\section{Conclusion}
We developed CdTe diode double-sided strip detectors that have Al anode
and Pt cathode strips. Two types of prototypes were assembled and we evaluated their 
imaging capability and spectral performance. Gamma-ray imaging spectroscopy with 
a position resolution of 400~$\mu$m was demonstrated by using the 2.6~cm CdTe DSD.
The energy resolution of 1.8~keV (FWHM) at 59.54~keV was obtained with the 1.3~cm CdTe DSD.

For semiconductor detector readout, we have developed a new analog ASIC, VA32TA6, characterized by 
a main new feature of including an on-chip ADC. By constructing CdTe pad detectors,
we tested the functions and performance of VA32TA6. The ADC worked 
properly and good noise performance was obtained.

We are developing the next version of the CdTe DSD, that will use
VA32TA6s as the readout ASICs. By using VA32TA6s,
the readout system may be simplified, because it is possible 
to control the ASICs and read data by only using digital signals.
Such a new configuration would offer improvements in performance as required
for the the ASTRO-H (NeXT) Hard X-ray Imager and future X-ray/gamma-ray telescopes.

\section*{Acknowledgment}
This work was supported by KAKENHI(19740172).


\ifCLASSOPTIONcaptionsoff
  \newpage
\fi

\end{document}